\documentclass[aps,pre,showpacs,amsmath,amssymb,reprint]{revtex4-1}

\usepackage{bm}
\usepackage{graphicx}

\begin{document}

\title{
Single-wavenumber Representation of Nonlinear Energy Spectrum\\ 
in Elastic-Wave Turbulence of {F}\"oppl-von {K}\'arm\'an Equation:\\
Energy Decomposition Analysis and Energy Budget
}

\author{Naoto Yokoyama}
\email{yokoyama@kuaero.kyoto-u.ac.jp}
\affiliation{Department of Aeronautics and Astronautics, Kyoto University, Kyoto 615-8540, Japan}

\author{Masanori Takaoka}
\email{mtakaoka@mail.doshisha.ac.jp}
\affiliation{Department of Mechanical Engineering, Doshisha University, Kyotanabe 610-0394, Japan}

\date{\today}

\begin{abstract}

A single-wavenumber representation of nonlinear energy spectrum, i.e., stretching energy spectrum
is found in elastic-wave turbulence
governed by the F\"{o}ppl-von K\'{a}rm\'{a}n (FvK) equation.
The representation enables energy decomposition analysis
in the wavenumber space,
and analytical expressions of detailed energy budget in the nonlinear interactions
are obtained for the first time in wave turbulence systems.
We numerically solved the FvK equation and observed the following facts.
Kinetic and bending energies are comparable with each other at large wavenumbers
as the weak turbulence theory suggests.
On the other hand,
the stretching energy is larger than the bending energy
at small wavenumbers,
i.e., the nonlinearity is relatively strong.
The strong correlation between a mode $a_{\bm{k}}$ 
and its companion mode $a_{-\bm{k}}$
is observed at the small wavenumbers.
Energy transfer shows that 
the energy is input into the wave field through stretching-energy transfer
at the small wavenumbers, and dissipated through the quartic part of 
kinetic-energy transfer at the large wavenumbers.
A total-energy flux consistent with the energy conservation
is calculated directly by using the analytical expression of the total-energy transfer,
and the forward energy cascade is observed clearly.

\end{abstract}


\maketitle

\section{Introduction}

Energy decomposition analysis helps understanding of 
the mechanism of energy distribution.
Exchange between kinetic energy and potential energy
is observed in oscillatory or wave motion,
while the total energy is conserved.
The exchange is seen as elliptic motion,
which can be distorted by the nonlinearity, in the phase space.
In Refs.~\cite{llelasticity,audoly2010elasticity},
the energy is decomposed into kinetic, bending and stretching energies
to derive the governing equation of the elastic waves.

In the relaxation, known as selective decay process, 
of hydrodynamic turbulent flows,
the depression of nonlinearity has been often discussed.
The strong correlations between velocity and vorticity
have been reported in hydrodynamic turbulence;
e.g., parallelization of velocity and vorticity called Beltramization
in three-dimensional flow~\cite{:/content/aip/journal/pof1/31/9/10.1063/1.866591,*PhysRevLett.54.2505,*:/content/aip/journal/pof1/30/8/10.1063/1.866513},
and negative temperature state such as the sinh-Poisson state
in two-dimensional flow~\cite{PhysRevLett.66.2731,*:/content/aip/journal/pof1/25/1/10.1063/1.863609,*:/content/aip/journal/pofa/4/1/10.1063/1.858525}.
These relaxed states have the correlation among modes.
It is in contrast with the weak turbulence,
where the independence among modes are presupposed.
In the Majda-McLaughlin-Tabak (MMT) model,
which is a one-dimensional mathematical model of wave turbulence,
spatially-localized coherent structures are reported~\cite{Cai2001551}.
\citeauthor{zakharov2001}~\cite{zakharov2001}
modified MMT model to fit the weak turbulence theory (WTT) 
by introducing a nonlinear term
that prevents the correlation of modes.
We will here report correlations between each pair of modes at large scales
in elastic-wave turbulence,
which is consistent with our previous work
where the separation wavenumber between the weak and strong turbulence
is estimated via the applicability limit of the random phase approximation (RPA) in WTT~\cite{PhysRevE.89.012909}.

Since the coexistence of {\em non-weak\/} and {\em weak\/} turbulence will be investigated
in this paper,
we here distinguish {\em wave\/} turbulence and {\em weak\/} turbulence:
the former is referred to as a wave turbulent state
where the nonlinear interactions are not necessarily weak,
and the latter is a wave turbulent state where WTT can be applied.
Thus, {\em wave\/} turbulence includes {\em weak\/} turbulence.

Fourier spectral representation is widely used
in the analysis of the homogeneous turbulence governed by the Navier-Stokes equation,
because one of the most important study objectives is
to clarify energy distribution
formed by hierarchical structures over a wide range of scales.
The so-called cascade theory, 
which was proposed by Kolmogorov~\cite{K41a}
as the first statistical theory of turbulence,
predicts the direction of energy transfer 
and is well described in the wavenumber space.
Also in researches of the weak turbulence systems,
the Fourier spectral representation is convenient
to introduce the complex amplitudes as elementary waves 
to apply RPA in WTT.

The analysis of the wave turbulence is confronted with the following difficulties,
which stem from the fact that
only the quadratic quantities of the complex amplitudes
have been considered as energy.
More properly, the quadratic energy corresponds to the linear part of the dynamics
and the ensemble-averaged quadratic energy is conserved only in the weakly nonlinear limit,
even if its dynamics is governed by a Hamiltonian.
Although it is convenient to use the complex amplitudes
in application of RPA to derive the kinetic equation,
the perturbative expansion of the complex amplitudes is inevitable 
to represent the nonlinearity of the system.
The nonlinear energy appears as convolutions of the complex amplitudes,
since the complex amplitudes are introduced for the different purpose.
On the other hand, for example, in the Navier-Stokes turbulence
the energy is given by a single-wavenumber representation like $|\bm{u}_{\bm{k}}|^2/2$,
and this kind of problem does not appear,
since the energy is simply given by the quadratic form by its nature.

To analyze energy budget,
it is indispensable to take into account 
the full Hamiltonian dynamics.
A single-wavenumber representation of the higher-order energy is required
to identify the nonlinear dynamics at each scale.
In addition to the nonconservation of the quadratic energy,
its transfer in the wavenumber space 
cannot be obtained as a closed expression in the representation of the complex amplitudes.
If a single-wavenumber representation of the energy can be found,
the explicit expression of the detailed energy budget is obtained,
even {\em not\/} in the weakly nonlinear limit.

Demanding the constancy of the energy flux and the complete self-similarity,
the dimensional analysis using a specific form of the kinetic equation
predicts the Kolmogorov-Zakharov spectrum
as described in Chap.~3 of Ref.~\cite{zak_book}.
While the spectral form can be obtained easily in this approach,
the Kolmogorov-Zakharov spectrum can be obtained also
as a stationary solution of the kinetic equation
with help from the so-called Zakharov transformation.
The energy flux in the framework of WTT
can be represented by the collision integral.
Care should be taken to distinguish WTT and the Hamiltonian dynamics,
since the ensemble-averaged quadratic energy is conserved only under the kinetic equation.
Although the quadratic-energy fluxes for a variety of spectral parameters
were numerically obtained in Ref.~\cite{resio},
no total-energy flux
has been obtained so far even in the weakly nonlinear limit.
We will here report that 
the flux of the {\em total\/} energy is directly calculated for the first time by using the analytical expression 
for the transfer.

The energy flux {\em not\/} in the weakly nonlinear limit
is difficult to obtain.
The most primitive estimation of the energy flux $\mathcal{P}(k)$ through $k=|\bm{k}|$
is obtained from the cumulative energy $\widetilde{\mathcal{E}}(k)$, 
the cumulative energy input $\widetilde{\mathcal{F}}(k)$,
and the cumulative energy dissipation $\widetilde{\mathcal{D}}(k)$ between $0$ and $k$
by using the scale-by-scale energy budget equation 
$\mathcal{P}(k) = -\partial \widetilde{\mathcal{E}}(k)/ \partial t
+ \widetilde{\mathcal{F}}(k) - \widetilde{\mathcal{D}}(k)$~\cite{frisch1995turbulence}.
The energy flux in a statistically-steady state is usually estimated by measuring the energy
injected into the system
when the dissipation is localized at large wavenumbers~\cite{boudaoud2008observation,*mordant2008there}.
The energy flux obtained in Ref.~\cite{PhysRevE.89.062925},
which is defined as $\widetilde{\mathcal{F}}(k) - \widetilde{\mathcal{D}}(k)$,
is the same as the flux estimated only by the energy input
for the dissipation localized at the large wavenumbers.
Their approaches do not contain the expression derived from the nonlinear term of the governing equation.
The constancy in the inertial subrange
of the energy flux estimated from $\widetilde{\mathcal{F}}(k) - \widetilde{\mathcal{D}}(k)$ is an obvious consequence from
the localization of the external force and dissipation,
and the constancy is independent of whether the nonlinear interactions are local or not.
The statistical steadiness $\partial \widetilde{\mathcal{E}}(k) / \partial t=0$
should be rigorously verified.
Furthermore,
the energy injected into the system is not necessarily in strict accordance with
the energy flux that cascades in the inertial subrange~\cite{0295-5075-102-3-30002}.
In laboratory experiments of surface waves,
the energy flux is estimated indirectly
by the energy decay rate after switching off the energy input
or by the dissipation spectrum.
This estimation requires additional assumptions,
because it is the power spectrum of the displacement 
that can be obtained experimentally~\cite{PhysRevLett.99.014501,*PhysRevE.89.023003}. 
The energy flux may be evaluated by using 
structure functions in the real space,
though it is a little different from that defined in the wavenumber space.
Even in direct numerical simulations according to dynamical equations,
the energy flux consistent with the energy conservation has not been obtained directly~\cite{Rumpf2005188,pushkarev,*zakharov_steady,*PhysRevLett.96.204501}.

The elastic-wave turbulence, 
which is tractable experimentally, numerically and theoretically,
exhibits rich phenomena:
weak turbulence~\cite{during2006weak,boudaoud2008observation,*mordant2008there},
spatio-temporal dynamics~\cite{PhysRevLett.103.204301,*mordant2010fourier},
spectral variation~\cite{PhysRevLett.107.034501,PhysRevE.89.012909}
and strongly nonlinear structures~\cite{PhysRevLett.111.054302}.
Among them,
the coexistence of the weakly nonlinear spectrum 
and a strongly nonlinear spectrum is one of the most remarkable properties~\cite{PhysRevLett.110.105501,PhysRevE.89.012909}.
It is an interesting challenge to clarify the energy budget
in the state where the weak turbulence and the strong turbulence coexist.
It should be noted here that
we use ``strong'' as short-hand notation 
to represent the relatively strongly nonlinear state
whose nonlinearity is {\em not\/} so strong as to break the first-principle dynamical equations,
but as strong as to break the weak nonlinearity assumption in WTT.

In this paper, 
we analyze the wave turbulence in a thin elastic plate
by numerical simulations according to the F\"{o}ppl-von K\'{a}rm\'{a}n equation.
The single-wavenumber representation of the nonlinear energy spectrum
opens a way for the above difficulties.
It enables the energy decomposition analysis and
the investigations of the energy budget due to the nonlinear interactions.
The next section is devoted to the formulation of the problem
with focusing on the Fourier representation of the system.
In Sec.~\ref{sec:results}, two kinds of numerical results are shown.
One is the energy decomposition analysis,
and the other is the energy budget.
The last section is devoted to concluding remark.

\section{Formulation}
\label{sec:formulation}

\subsection{Governing equation and numerical scheme}

The dynamics of elastic waves propagating in a thin plate
is described by the F\"{o}ppl-von K\'{a}rm\'{a}n (FvK) equation
for the displacement $\zeta$ and the momentum $p$
via the Airy stress potential $\chi$~\cite{llelasticity,audoly2010elasticity}.
Under the periodic boundary condition,
the FvK equation is written as
\begin{subequations}
\begin{align}
&
\frac{d\zeta_{\bm{k}}}{dt} = \frac{p_{\bm{k}}}{\rho}
,
\quad
\frac{dp_{\bm{k}}}{dt}  = - \rho \omega_{\bm{k}}^2 \zeta_{\bm{k}}
 + \!\!\!\!
\sum_{\bm{k}_1+\bm{k}_2=\bm{k}}
\!\!\!\!
 |\bm{k}_1 \times \bm{k}_2|^2 \zeta_{\bm{k}_1} \chi_{\bm{k}_2}
,
\\
&
\chi_{\bm{k}} = -\frac{Y}{2k^4} \sum_{\bm{k}_1+\bm{k}_2=\bm{k}} |\bm{k}_1 \times \bm{k}_2|^2 \zeta_{\bm{k}_1} \zeta_{\bm{k}_2}
,
\label{eq:chik}%
\end{align}%
\label{eq:FvKink}%
\end{subequations}%
where $\zeta_{\bm{k}}$, $p_{\bm{k}}$ and $\chi_{\bm{k}}$
are the Fourier coefficients of the displacement, of the momentum, and of the Airy stress potential, respectively.
The Young's modulus $Y$ and the density $\rho$
are the material quantities of an elastic plate.
The frequency $\omega_{\bm{k}}$
is given by the linear dispersion relation:
\begin{align}
 \omega_{\bm{k}} = \sqrt{\frac{Yh^2}{12(1-\sigma^2)\rho}} \ k^2
,
\label{eq:lineardispersion}
\end{align}
where $\sigma$ and $h$ are respectively the Poisson ratio and the thickness of the elastic plate.

The complex amplitude is defined as
\begin{align}
 a_{\bm{k}} = 
\frac{\rho \omega_{\bm{k}} \zeta_{\bm{k}} + i p_{\bm{k}}}{\sqrt{2 \rho \omega_{\bm{k}}}}
.
\label{eq:ComplexAmplitude}
\end{align}
The complex amplitude is used as the elementary wave of the wavenumber $\bm{k}$ in WTT.
Then,
the variables in Eq.~(\ref{eq:FvKink})
are given as
\begin{subequations}
\begin{align}
 \zeta_{\bm{k}} &= \frac{1}{\sqrt{2\rho \omega_{\bm{k}}}} (a_{\bm{k}} + a_{-\bm{k}}^{\ast})
,\label{eq:a2zeta}
\\
 p_{\bm{k}} &= -i \sqrt{\frac{\rho \omega_{\bm{k}}}{2}} (a_{\bm{k}} - a_{-\bm{k}}^{\ast})
,
\\
 \chi_{\bm{k}} &= - \frac{Y}{4\rho k^4}
\!\!\!\!
 \sum_{\bm{k}_1+\bm{k}_2=\bm{k}}
\!\!\!\!
 \frac{|\bm{k}_1 \times \bm{k}_2|^2}{\sqrt{\omega_{\bm{k}_1} \omega_{\bm{k}_2}}}
(a_{\bm{k}_1} + a_{-\bm{k}_1}^{\ast})
(a_{\bm{k}_2} + a_{-\bm{k}_2}^{\ast})
,
\end{align}%
\end{subequations}%
where $a^{\ast}$ represents the complex conjugate of $a$.
Equation~(\ref{eq:FvKink}) is reduced to a single equation 
for $a_{\bm{k}}$ as
\begin{align}
 \frac{da_{\bm{k}}}{dt} 
=& - i \omega_{\bm{k}} a_{\bm{k}}
\nonumber\\
&
- \frac{iY}{8\rho^2}
\!\!
 \sum_{\bm{k}_1+\bm{k}_2+\bm{k}_3=\bm{k}}
\!\!\!\!\!\!\!\!
\frac{|\bm{k} \times \bm{k}_1|^2|\bm{k}_2 \times \bm{k}_3|^2}{|\bm{k}_2+\bm{k}_3|^4 }
\nonumber\\
& \quad \times
\frac{
(a_{\bm{k}_1} + a_{-\bm{k}_1}^{\ast})
(a_{\bm{k}_2} + a_{-\bm{k}_2}^{\ast})
(a_{\bm{k}_3} + a_{-\bm{k}_3}^{\ast})
}{\sqrt{\omega_{\bm{k}}\omega_{\bm{k}_1}\omega_{\bm{k}_2}\omega_{\bm{k}_3}}}
.
\label{eq:fvka}
\end{align}
The first term in the right-hand side corresponds to the linear harmonic oscillation,
and the second one to the nonlinear interactions.

Direct numerical simulations (DNS) according to Eq.~(\ref{eq:fvka})
are performed with the parameter values as
$\rho=7.8 \times 10^3$kg/m$^3$, $Y = 2.0 \times 10^{11}$Pa, $\sigma = 0.30$,
and $h = 5.0 \times 10^{-4}$m.
The plate is supposed to have the periodic boundary of $1$m$\times 1$m.
Thus,
the two-dimensional wavenumber vector $\bm{k}$ is discretized as $\bm{k} \in (2\pi \mathbb{Z})^2$.
The pseudo-spectral method is employed 
and
the number of the aliasing-free modes is $512 \times 512$.
Since the 4/2-law is required to remove the aliasing errors in the third-order nonlinearity,
we use $1024 \times 1024$ mode in the calculation of the convolutions.

The external force $F_{\bm{k}}$
and the dissipation $D_{\bm{k}}$
are added to the right-hand side of Eq.~(\ref{eq:fvka})
to make statistically-steady non-equilibrium states.
The external force $F_{\bm{k}}$ are added
so that $a_{\bm{k}}$'s at the small wavenumbers $|\bm{k}| \leq 8\pi$
have a magnitude constant in time,
while the phases of $a_{\bm{k}}$'s are determined by Eq.~(\ref{eq:fvka}).
The dissipation is added as $D_{\bm{k}} = -\nu |\bm{k}|^8 a_{\bm{k}}$,
where $\nu = 1.21 \times 10^{-22}$.
As we can recognize from Figs.~\ref{fig:alles} and \ref{fig:transfer},
which appear below,
the dissipation is effective in the wavenumber range
$|\bm{k}| \gtrapprox 256\pi$.
The exponential decay of the energy spectra shown in Fig.~\ref{fig:alles}
at the large wavenumbers
gives the assurance of our DNS with this mode number.
Details of the numerical scheme 
are explained in Ref.~\cite{PhysRevLett.110.105501}.

It is preferable for the external force and the dissipation to be localized in scales
to achieve a large inertial subrange of turbulence spectra.
Although it is reported that
broadly-affecting Lorentzian dissipation successfully
reproduces the experimentally-observed spectrum~\cite{PhysRevLett.111.054302},
we are interested in the properties in the inertial subrange 
in the FvK turbulence.
According to the derivation of the equation,
it might be realized and examined in laboratory,
if one could perform the experiment in the vacuum environment
to reduce drags acting on the thin plate, e.g., induced mass,
by using much less dissipative plates to reduce internal friction.

\subsection{Hamiltonian and energy decomposition}
\label{ssec:decomposition}

The FvK equation~(\ref{eq:FvKink}) can be written as a canonical equation:
\begin{align*}
\frac{d \zeta_{\bm{k}}}{dt} = \frac{\delta \mathcal{H}}{\delta p^{\ast}_{\bm{k}}}
,
\quad
\frac{d p_{\bm{k}}}{dt} = - \frac{\delta \mathcal{H}}{\delta \zeta^{\ast}_{\bm{k}}}
,
\end{align*}%
when we introduce the Hamiltonian $\mathcal{H}$ as
\begin{align}
 \mathcal{H} =& \sum_{\bm{k}} \left(\frac{1}{2\rho} |p_{\bm{k}}|^2 + \frac{\rho \omega_{\bm{k}}^2}{2} |\zeta_{\bm{k}}|^2 \right)
\nonumber\\
& 
+\frac{Y}{8}
\!\!\!
 \sum_{\bm{k}+\bm{k}_1-\bm{k}_2-\bm{k}_3=\bm{0}}
\!\!\!\!\!\!\!\!\!\!\!
\frac{|\bm{k} \times \bm{k}_1|^2 |\bm{k}_2 \times \bm{k}_3|^2}{|\bm{k}_2 + \bm{k}_3|^4}
\zeta_{\bm{k}}^{\ast} \zeta_{\bm{k}_1}^{\ast} \zeta_{\bm{k}_2} \zeta_{\bm{k}_3}
,
\label{eq:hamiltonian}
\end{align}
where $\delta/\delta \zeta^{\ast}_{\bm{k}}$ and $\delta/\delta p^{\ast}_{\bm{k}}$ express 
the functional derivatives
with respect to $\zeta^{\ast}_{\bm{k}}$ and $p^{\ast}_{\bm{k}}$, respectively.
Use has been made of $\zeta_{\bm{k}} = \zeta_{-\bm{k}}^{\ast}$
to rewrite the second term in right-hand side into the symmetric form.
Note that $\zeta_{\bm{k}}$ ($p_{\bm{k}}$) and $\zeta_{\bm{k}}^{\ast}$ ($p_{\bm{k}}^{\ast}$) are not independent
of each other.
The relation to the conventional representation with the complex amplitudes
in WTT is given in Appendix~\ref{sec:Appendix}.

The Hamiltonian consists of three kinds of energies, i.e.,
the kinetic energy, the bending energy, and the stretching energy~\cite{audoly2010elasticity}.
The bending energy derives from the out-of-plane displacement,
while the stretching energy comes from the in-plane strain.

The total energy of each mode $E_{\bm{k}}$ is
the sum of the kinetic energy $K_{\bm{k}}$
and the potential energy $V_{\bm{k}}$,
i.e., $E_{\bm{k}} = K_{\bm{k}} + V_{\bm{k}}$.
The potential energy of each mode
is the sum of the bending energy $V_{\mathrm{b} \bm{k}}$ and 
the stretching energy $V_{\mathrm{s} \bm{k}}$, i.e.,
$V_{\bm{k}} = V_{\mathrm{b} \bm{k}} + V_{\mathrm{s} \bm{k}}$.
Here,
\begin{subequations}
\begin{align}
 K_{\bm{k}} &= \frac{1}{2\rho} |p_{\bm{k}}|^2
 = \frac{\omega_{\bm{k}}}{4}  \left( |a_{\bm{k}}|^2 + |a_{-\bm{k}}|^2 - 2\mathrm{Re}(a_{\bm{k}}a_{-\bm{k}}) \right)
 ,
 \label{eq:defenergyiesK}
 \\
 V_{\mathrm{b} \bm{k}} &= \frac{\rho \omega_{\bm{k}}^2}{2} |\zeta_{\bm{k}}|^2
 = \frac{\omega_{\bm{k}}}{4} \left( |a_{\bm{k}}|^2 + |a_{-\bm{k}}|^2 + 2\mathrm{Re}(a_{\bm{k}}a_{-\bm{k}}) \right)
 ,
 \label{eq:defenergyiesVb}
 \\
 V_{\mathrm{s} \bm{k}} &= \frac{k^4}{2Y} |\chi_{\bm{k}}|^2
 = \frac{Y}{32 \rho^2 k^4}
 \!\!\!\!
 \sum_{\substack{\bm{k}_1+\bm{k}_2=\bm{k} \\ \bm{k}_3+\bm{k}_4=\bm{k}}}
 \!\!\!\!
 \frac{|\bm{k}_1 \times \bm{k}_2|^2 |\bm{k}_3 \times \bm{k}_4|^2}{\sqrt{\omega_{\bm{k}_1} \omega_{\bm{k}_2}\omega_{\bm{k}_3} \omega_{\bm{k}_4}}}
 \nonumber\\
 &
 \!\!\!\!
 \times
 (a_{\bm{k}_1}^{\ast} + a_{-\bm{k}_1})
 (a_{\bm{k}_2}^{\ast} + a_{-\bm{k}_2})
 (a_{\bm{k}_3} + a_{-\bm{k}_3}^{\ast})
 (a_{\bm{k}_4} + a_{-\bm{k}_4}^{\ast})
 .
 \label{eq:defenergyiesVs}
\end{align}%
\label{eq:defenergyies}%
\end{subequations}
The quadratic energy of each mode
is given as the sum of the kinetic energy and the bending energy,
i.e.,
$E^{(2)}_{\bm{k}} = K_{\bm{k}} + V_{\mathrm{b} \bm{k}}$,
because both energies are $O(|a|^2)$.
On the other hand,
the quartic energy $E^{(4)}_{\bm{k}}$ is the stretching energy $V_{\mathrm{s} \bm{k}}$,
which is $O(|a|^4)$.
The Hamiltonian~(\ref{eq:hamiltonian}) can also be written in terms of these energies as
\begin{align}
\mathcal{H}=\sum_{\bm{k}} E_{\bm{k}} =\sum_{\bm{k}} (E^{(2)}_{\bm{k}}+ E^{(4)}_{\bm{k}} )
= \sum_{\bm{k}} ( K_{\bm{k}} + V_{\mathrm{b} \bm{k}} + V_{\mathrm{s} \bm{k}} )
.
\label{eq:HamiltonianK-sum}
\end{align}

It should be emphasized that
usage of the Fourier coefficient of the Airy stress potential $\chi_{\bm{k}}$,
given as Eq.~(\ref{eq:chik}) 
enables the representation of the nonlinear energy for a single-wavenumber mode as Eq.~(\ref{eq:defenergyiesVs})
in this system.
The complex amplitude $a_{\bm{k}}$ is introduced as the elementary wave in WTT.
When the system's Hamiltonian is expanded in terms of $a_{\bm{k}}$, 
it leads to the nonlinear energy in the form of a convolution consisting of the four wavenumbers
as shown in Eq.~(\ref{eq:defenergyiesVs}).
We here consider $\zeta_{\bm{k}}$, $p_{\bm{k}}$ and $\chi_{\bm{k}}$ as elementary waves
in the representation of the energies, $K_{\bm{k}}$, $V_{\mathrm{b} \bm{k}}$ and $V_{\mathrm{s} \bm{k}}$.

In the framework of WTT,
the energy of $\bm{k}$ is defined as the quadratic energy:
$E_{\bm{k}}^{\mathrm{WTT}} = \omega_{\bm{k}} \langle |a_{\bm{k}}|^2 \rangle$,
where $\langle \cdot \rangle$ denotes the ensemble averaging.
The quadratic energy in our notation
and the energy in WTT
are related as $\langle E^{(2)}_{\bm{k}} \rangle = \langle K_{\bm{k}} + V_{\mathrm{b} \bm{k}} \rangle = E_{\bm{k}}^{\mathrm{WTT}}+E_{-\bm{k}}^{\mathrm{WTT}}$.
The energy in WTT $\sum_{\bm{k}} E_{\bm{k}}^{\mathrm{WTT}}$ is not conserved under the FvK equation
regardless of the ensemble-averaging,
since it lacks the stretching energy $V_{\mathrm{s} \bm{k}}$
in the Hamiltonian (\ref{eq:HamiltonianK-sum}),
i.e., $\sum_{\bm{k}} E_{\bm{k}}^{\mathrm{WTT}} = \sum_{\bm{k}} \langle E_{\bm{k}}^{(2)} \rangle = \langle \mathcal{H}_2 \rangle \neq \mathcal{H}$,
where $\mathcal{H}_2$ represents the quadratic part of the Hamiltonian.
It should be noted that
$E_{-\bm{k}}^{\mathrm{WTT}}$ is independent of $E_{\bm{k}}^{\mathrm{WTT}}$,
but $E_{\bm{k}}^{(2)} = E_{-\bm{k}}^{(2)}$ as well as $E_{\bm{k}} = E_{-\bm{k}}$, 
because $E_{\bm{k}}^{(2)}$ and $E_{\bm{k}}$ are given by the Fourier coefficients of the real-valued functions.

\section{Results}
\label{sec:results}

We will show the numerical results for the moderate energy level,
which corresponds to EL3 in Ref.~\cite{PhysRevE.89.012909}.
This energy level is chosen so as to realize the coexistence 
of the weak and strong energy spectra.
The number of the modes are twice those in Ref.~\cite{PhysRevE.89.012909} in each direction
to obtain larger inertial subrange.
The FvK equation is applicable for this energy level,
because the root mean square of the gradient of the displacement 
$\langle |\nabla \zeta|^2 \rangle^{1/2} \approx 0.15$~\footnote{
The average is performed over $1024 \times 4 \times 512^2$ points:
$1024$ independent realizations,
$4$ different times at an interval sufficiently longer than the longest linear period,
and $512^2$ grid points.
}.
Furthermore, this energy level looks intermediate between
the two fields reported in Fig.~2 of Ref.~\cite{PhysRevLett.111.054302},
i.e., much smaller than the energy level 
at which the dynamic crumpling appears.

\subsection{Decomposed energy spectra and correlation between companion modes}

\begin{figure}
 \includegraphics[scale=1.15]{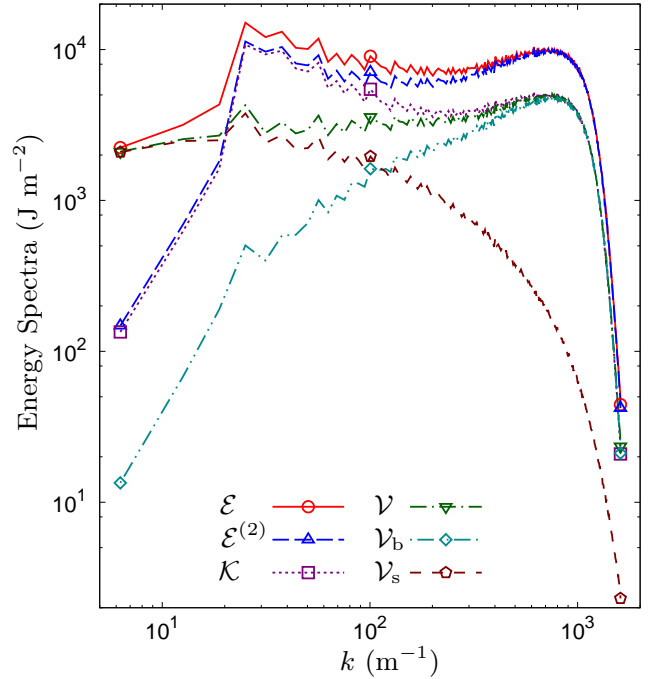}
 \caption{(Color online)
Energy spectra
 of the total energy $\mathcal{E}$, 
the quadratic energy $\mathcal{E}^{(2)}$,
the kinetic energy $\mathcal{K}$, the potential energy $\mathcal{V}$,
the bending energy $\mathcal{V}_{\mathrm{b}}$, 
and the stretching energy $\mathcal{V}_{\mathrm{s}}$.
 \label{fig:alles}
 }
\end{figure}
The azimuthally-integrated energy spectra
of the total energy $\mathcal{E}(k)$, 
the quadratic energy $\mathcal{E}^{(2)}(k)$,
the kinetic energy $\mathcal{K}(k)$, 
the potential energy $\mathcal{V}(k)$,
the bending energy $\mathcal{V}_{\mathrm{b}}(k)$, and the stretching energy $\mathcal{V}_{\mathrm{s}}(k)$
are shown in Fig.~\ref{fig:alles}.
The azimuthally-integrated spectrum of the total energy,
for example,
is defined as
$\mathcal{E}(k) = (\Delta k)^{-1} \sum_{k-\Delta k/2 \leq |\bm{k}^{\prime}| < k+\Delta k/2} \langle E_{\bm{k}^{\prime}} \rangle$,
where $\Delta k$ is the width of the bins to make the azimuthal integration~\footnote{
The spectra are obtained by averaging over 
$4096=1024\times 4$ fields:
$1024$ independent realizations which are started from different initial conditions,
and $4$ different times at an interval sufficiently longer than the longest linear period.}.
(See also Appendix in Ref.~\cite{PhysRevE.89.012909}.)
Note that the azimuthally-integrated spectrum of the energy in WTT
is equal to that of the quadratic energy, i.e., $\mathcal{E}^{\mathrm{WTT}}(k) = \mathcal{E}^{(2)}(k)$
because of the statistical isotropy.

In Ref.~\cite{PhysRevLett.110.105501},
the quadratic energy $\mathcal{E}^{(2)}(k)$ was examined to compare with WTT,
and the coexistence of the weakly and strongly nonlinear energy spectra was found.
It was also found that
the coexistence in $\mathcal{E}^{(2)}(k)$
results from the coexistence in the kinetic energy $\mathcal{K}(k)$~\cite{PhysRevE.89.012909}.
The coexistence is observed also in Fig.~\ref{fig:alles}:
the weakly nonlinear spectrum $\mathcal{E}^{(2)}(k) \propto k$ in the large wavenumbers,
and a strongly nonlinear spectrum $\mathcal{E}^{(2)}(k) \propto k^{-1/3}$ in the small wavenumbers.
The weakly nonlinear spectrum is a stationary solution of the kinetic equation~\cite{during2006weak}.
The strongly nonlinear spectrum is shallower than that observed in Ref.~\cite{PhysRevLett.111.054302}.
The difference between the strongly nonlinear spectra should be
caused by the difference between the external forces and between the dissipations.
The flexion of $\mathcal{K}(k)$
is in contrast with the monotonic increase of the bending energy $\mathcal{V}_{\mathrm{b}}(k)$.
As explained below,
this is due to an opposite effect
of the strong correlation between the pairs of modes 
caused by the nonlinear term.

At the large wavenumbers,
the kinetic and bending energies,
$\mathcal{K}(k)$ and $\mathcal{V}_{\mathrm{b}}(k)$, are comparable with each other.
It corresponds to the fact that
the average of the kinetic energy is equal to that of the potential energy
in linear harmonic waves.
Because the quartic energy $\mathcal{V}_{\mathrm{s}}(k)$
is much smaller than the quadratic energy $\mathcal{E}^{(2)}(k)$,
the weak nonlinearity and the randomness of the phases at the large wavenumbers are confirmed.
(See the blue long-dashed curve and the brown short-dashed curve in Fig.~\ref{fig:alles}.)
Therefore, WTT works well in this wavenumber range.

At the small wavenumbers, in contrast, 
$\mathcal{V}_{\mathrm{s}}(k)$ is larger than $\mathcal{V}_{\mathrm{b}}(k)$.
Therefore, the nonlinearity is relatively strong there.
(In this paper, we simply refer to it as strong nonlinearity.)
The kinetic energy
$\mathcal{K}(k)$ accounts for most of the total energy $\mathcal{E}(k)$,
and $\mathcal{E}^{(2)}(k)$ is larger than $\mathcal{V}_{\mathrm{s}}(k)$
even at these small wavenumbers.
The non-smallness of $\mathcal{V}_{\mathrm{s}}(k)$ at the small wavenumbers,
especially at $k\leq 8\pi$, 
stems from the nonlocality of the nonlinear term in the wavenumber space,
as is known from the fact that $\mathcal{V}_{\mathrm{s}}(k)$ is obtained 
via the convolution (\ref{eq:defenergyiesVs}).
The kinetic energy, furthermore, is closely related to the stretching energy
via the energy transfer as shown in the next subsection.

\begin{figure}
 \includegraphics[scale=1.15]{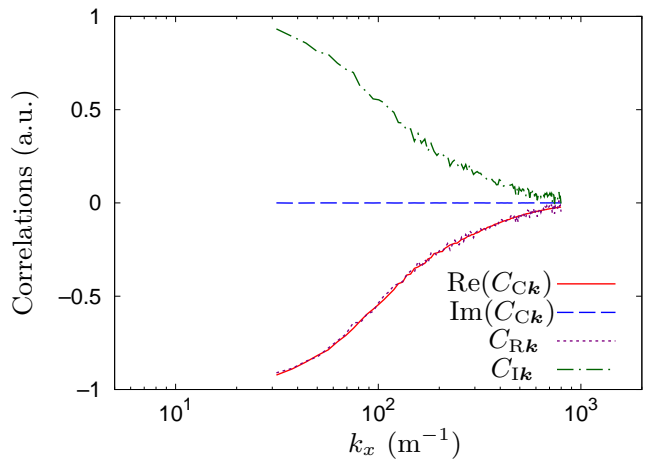} 
 \caption{(Color online)
 Correlation between $a_{\bm{k}}$ and $a_{-\bm{k}}$.
 ($\bm{k} = (k_x,0)$ and $k_x \in [10\pi, 256\pi]$)
\label{fig:correlation}}
\end{figure}

The deviation of the kinetic energy $\mathcal{K}(k)$
from the bending energy $\mathcal{V}_{\mathrm{b}}(k)$ at the small wavenumbers
comes from the term $\mathrm{Re}(a_{\bm{k}}a_{-\bm{k}})$,
which is found as the difference between Eqs.~(\ref{eq:defenergyiesK}) and (\ref{eq:defenergyiesVb}).
The correlation of the complex amplitudes between the companion modes,
$a_{\bm{k}}$ and $a_{-\bm{k}}$, 
is defined as
\begin{align}
 C_{\mathrm{C} \bm{k}}
 = \frac{\langle a_{\bm{k}} a_{-\bm{k}} \rangle}%
 {\sqrt{\langle |a_{\bm{k}}|^2 \rangle \langle |a_{-\bm{k}}|^2 \rangle}}
.
\end{align}
The independency between the complex amplitudes at the first order,
i.e., $\langle a_{\bm{k}} a_{\bm{k}^{\prime}}\rangle = 0$
is required by RPA in WTT.
Similarly,
the correlations of the real and imaginary parts of the companion modes
are defined as
\begin{subequations}
\begin{align}
 C_{\mathrm{R} \bm{k}}
 &= \frac{\langle \mathrm{Re}(a_{\bm{k}}) \mathrm{Re}(a_{-\bm{k}}) \rangle}%
 {\sqrt{\langle (\mathrm{Re}(a_{\bm{k}}))^2 \rangle \langle (\mathrm{Re}(a_{-\bm{k}}))^2 \rangle}}
,
\\
 C_{\mathrm{I} \bm{k}}
 &= \frac{\langle \mathrm{Im}(a_{\bm{k}}) \mathrm{Im}(a_{-\bm{k}}) \rangle}%
 {\sqrt{\langle (\mathrm{Im}(a_{\bm{k}}))^2 \rangle \langle (\mathrm{Im}(a_{-\bm{k}}))^2 \rangle}}
.
\end{align}
\end{subequations}

In Fig.~\ref{fig:correlation},
the correlations between companion modes 
at $\bm{k} = (k_x,0)$ and $-\bm{k} = (-k_x,0)$,
i.e., $C_{\mathrm{C} (k_x,0)}$, $C_{\mathrm{R} (k_x,0)}$, and $C_{\mathrm{I} (k_x,0)}$
are drawn in the range $k_x \in [10\pi, 256\pi]$
to avoid the influence from the artificially-added external force and dissipation.

At the large wavenumbers, where the nonlinearity is weak,
the correlations, $C_{\mathrm{C} \bm{k}}, C_{\mathrm{R} \bm{k}}$ and $C_{\mathrm{I} \bm{k}}$,
are almost zero.
It is consistent with RPA.
At the small wavenumbers, where the nonlinearity is relatively strong,
$C_{\mathrm{C} \bm{k}} \approx -1$,
$C_{\mathrm{R} \bm{k}} \approx -1$ and $C_{\mathrm{I} \bm{k}} \approx 1$.
It indicates $a_{\bm{k}} \approx -a_{-\bm{k}}^{\ast}$,
which is confirmed by the time series of $a_{\bm{k}}$ and $a_{-\bm{k}}$,
though the graphs are omitted here.
This fact is consistent with the results in Ref.~\cite{PhysRevE.89.012909},
where it is shown that the separation wavenumber which forms the division 
between the weakly and strongly nonlinear spectra
agrees with the critical wavenumber 
at which the nonlinear frequency shift is comparable with the linear frequency.
Namely, it means that RPA, which is the basis of WTT,
becomes inapplicable below the vicinity of the separation wavenumber.

In all the wavenumbers,
$\mathrm{Re}(C_{\mathrm{C} \bm{k}}) \approx C_{\mathrm{R} \bm{k}} \approx - C_{\mathrm{I} \bm{k}}$.
The curve for $\mathrm{Re}(C_{\mathrm{C} \bm{k}}$) is smoother than $C_{\mathrm{R} \bm{k}}$ and $C_{\mathrm{I} \bm{k}}$, 
since the former consists of the latter two elements,
i.e., the twice ensemble number.
If we decrease the amplitude of the external force,
the range of the wavenumbers where WTT holds becomes larger.
It is consistent with the results in Ref.~\cite{mordant2010fourier}.
The weak nonlinearity which results in
 $\langle a_{\bm{k}} a_{\bm{k}^{\prime}} \rangle = 0$ at the large wavenumbers
and the strongly nonlinear correlation $a_{\bm{k}} \approx -a_{-\bm{k}}^{\ast}$
at the small wavenumbers
make $\mathrm{Im}(C_{\mathrm{C} \bm{k}}) \approx 0$ over all the wavenumbers.

The strong correlation $a_{\bm{k}} \approx -a_{-\bm{k}}^{\ast}$
at the small wavenumbers
appears as $\mathcal{K}(k) \gg \mathcal{V}_{\mathrm{b}}(k)$ in Fig.~\ref{fig:alles},
which is consistent with Eqs.~(\ref{eq:defenergyiesK}) and (\ref{eq:defenergyiesVb}).
Because of Eq.~(\ref{eq:a2zeta}),
this correlation makes $\zeta_{\bm{k}}$ small.
It leads to depression
of the summand in the nonlinear term (see Eq.~(\ref{eq:fvka})),
which reminds us of the depression in the relaxation processes~\cite{:/content/aip/journal/pof1/31/9/10.1063/1.866591,*PhysRevLett.54.2505,*:/content/aip/journal/pof1/30/8/10.1063/1.866513,PhysRevLett.66.2731,*:/content/aip/journal/pof1/25/1/10.1063/1.863609,*:/content/aip/journal/pofa/4/1/10.1063/1.858525}
as written in the introduction.
It seems that this kind of the correlated states will survive
in contrast with the fast cascade of the uncorrelated modes.

One might think that this correlation, $a_{\bm{k}} \approx -a_{-\bm{k}}^{\ast}$,
contradicts to the strong nonlinearity
at the small wavenumbers,
since it appears to suppress the nonlinear term, 
the second term in the right-hand side of Eq.~(\ref{eq:fvka}).
The nonlinearity
can be large at the small wavenumbers
owing to the convolution,
which is the summation of the products of
$(a_{\bm{k}_1} + a_{-\bm{k}_1}^{\ast})$,
$(a_{\bm{k}_2} + a_{-\bm{k}_2}^{\ast})$,
and
$(a_{\bm{k}_3} + a_{-\bm{k}_3}^{\ast})$ at all wavenumbers,
because $(a_{\bm{k}_i} + a_{-\bm{k}_i}^{\ast})$ for $\bm{k}_i$ ($i=1, 2, 3$)
at the large wavenumbers are not small.
Namely, the nonlinearity at a wavenumber is not determined 
only by the elementary wave at the wavenumber.
This fact is also confirmed in Fig.~\ref{fig:alles}.
While the amplitudes of the linear energies, $\mathcal{E}^{(2)}$, $\mathcal{K}$ and $\mathcal{V}_{\mathrm{b}}$, 
decay at the small wavenumbers,
those including the nonlinear energy, $\mathcal{E}$, $\mathcal{V}$ and $\mathcal{V}_{\mathrm{s}}$, do not
and are almost constant $k \leq 8\pi$.

\subsection{Energy budget}

\begin{figure}
 \includegraphics[scale=1.05]{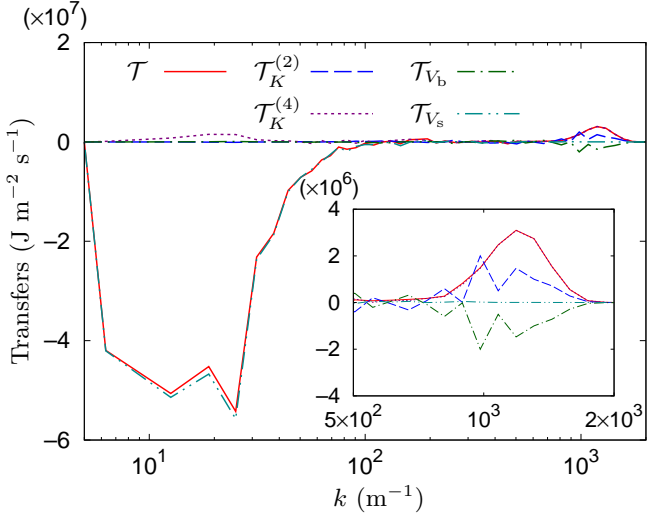}
 \caption{(Color online)
 Energy transfers of the total energy,
 of the quadratic and quartic parts of the kinetic energy,
 and of the bending and stretching energies.
 The abscissa is logarithmically scaled.
 The inset shows the enlargement at the large wavenumbers.
 \label{fig:transfer}
}
\end{figure}

To investigate the energy budget in detail,
our analysis here starts with energy transfer.
We define the energy transfer of $\bm{k}$
as $T_{\bm{k}} = \hat{d} E_{\bm{k}} /\hat{d}t$,
where the operator $\hat{d}/\hat{d}t$ expresses the time derivative
neglecting the external force and the dissipation.
According to the energy decomposition in Sec.~\ref{sec:formulation},
the total-energy transfer is also decomposed 
as
\begin{align}
 T_{\bm{k}} &=
\frac{\hat{d} K_{\bm{k}} }{\hat{d}t} + \frac{\hat{d} V_{b\bm{k}} }{\hat{d}t}
+ \frac{\hat{d} V_{s\bm{k}} }{\hat{d}t}
=
  T_{K \bm{k}}
  + T_{V_{\mathrm{b}} \bm{k}}
  + T_{V_{\mathrm{s}} \bm{k}}
 .
\end{align}
Corresponding to the linear and nonlinear terms in $dp_{\bm{k}}/dt$,
the transfer of the kinetic energy $T_{K \bm{k}}$ consists of
the quadratic and quartic parts,
$T_{K \bm{k}}^{(2)}$ and $T_{K \bm{k}}^{(4)}$,
i.e., $T_{K \bm{k}}=\hat{d} K_{\bm{k}} / \hat{d}t = T_{K \bm{k}}^{(2)}+T_{K \bm{k}}^{(4)}$.
From Eqs.~(\ref{eq:FvKink}) and (\ref{eq:defenergyies}),
\begin{subequations}
\begin{align}
T_{K \bm{k}}^{(2)}
&=
 - \frac{\omega_{\bm{k}}^2}{2} p_{\bm{k}}^{\ast} \zeta_{\bm{k}}
 + \mathrm{c.c.}
 ,
 \label{eq:transferK2}
 \\
 T_{K \bm{k}}^{(4)}
 &= 
 \frac{p_{\bm{k}}^{\ast}}{2\rho}
 \sum_{\bm{k}_1+\bm{k}_2=\bm{k}} \!\!\!\!
 |\bm{k}_1 \times \bm{k}_2|^2 \zeta_{\bm{k}_1} \chi_{\bm{k}_2}
 + \mathrm{c.c.}
 ,
 \label{eq:transferK4}
 \\
 T_{V_{\mathrm{b}} \bm{k}}
 &=
 \frac{\hat{d} V_{\mathrm{b} \bm{k}} }{\hat{d}t}
 =
 \frac{\omega_{\bm{k}}^2}{2} p_{\bm{k}}^{\ast} \zeta_{\bm{k}}
 + \mathrm{c.c.}
 ,
 \label{eq:transferVb}
 \\
T_{V_{\mathrm{s}} \bm{k}}
 &= 
 \frac{\hat{d} V_{\mathrm{s} \bm{k}} }{\hat{d}t}
 =
 - \frac{\chi_{\bm{k}}^{\ast}}{2\rho}
 \sum_{\bm{k}_1+\bm{k}_2=\bm{k}} \!\!\!\!
 |\bm{k}_1 \times \bm{k}_2|^2 p_{\bm{k}_1} \zeta_{\bm{k}_2} 
 + \mathrm{c.c.}
\label{eq:transferVs}
\end{align}%
\label{eq:energytransfers}%
\end{subequations}

Although the kinetic energy is represented as
a quadratic function of the complex amplitude,
its transfer has both quadratic part $T_{K \bm{k}}^{(2)}$
and quartic part $T_{K \bm{k}}^{(4)}$.
While the transfer of the bending energy $T_{V_{\mathrm{b}} \bm{k}}$ is a quadratic function of the complex amplitude,
that of the stretching energy $T_{V_{\mathrm{s}} \bm{k}}$ is a quartic function.

Apparently,
$T_{K \bm{k}}^{(2)}$ and $T_{V_{\mathrm{b}} \bm{k}}$ cancel each other,
representing the harmonic exchange between the kinetic and bending energies for a wavenumber.
Thus,
the quadratic parts of the transfer do not contribute
to the cascade between different scales.
In this sense,
to be exact,
$T_{K \bm{k}}^{(2)}$ and $T_{V_{\mathrm{b}} \bm{k}}$
are not transfers but transmutations from one form of the energy to the other.
Nonetheless,
we naively use the word ``transfers'' both for transfers and for transmutations.
The quartic-energy transfers,
$T_{K \bm{k}}^{(4)}$ and $T_{V_{\mathrm{s}} \bm{k}}$,
are the energy transfers
due to the nonlinear interactions among modes in the wavenumber space 
as known from Eqs.~(\ref{eq:transferK4}) and (\ref{eq:transferVs}).
They are of the same quartic order of the complex amplitude.
However,
only $T_{K \bm{k}}^{(4)}$ has been taken into account for the energy transfer in WTT
as $T_{\bm{k}}^{\mathrm{WTT}} = \hat{d} E_{\bm{k}}^{\mathrm{WTT}} /\hat{d}t$,
because it comes from the quadratic energy.
Namely,
$T_{\bm{k}}^{\mathrm{WTT}} + T_{-\bm{k}}^{\mathrm{WTT}} = \langle T_{K \bm{k}} + T_{V_{\mathrm{b}} \bm{k}} \rangle = \langle T_{K \bm{k}}^{(4)} \rangle$.

It should be emphasized that
the energy conservation holds only for the total energy,
which is the sum of the kinetic, bending and stretching energies,
but each decomposed energy is not conservative separately.
Namely,
$\sum_{\bm{k}} T_{\bm{k}} = 0$,
but
$\sum_{\bm{k}} T_{K \bm{k}}^{(2)}$, $\sum_{\bm{k}} T_{K \bm{k}}^{(4)}$, $\sum_{\bm{k}} T_{V_{\mathrm{b}} \bm{k}}$, $\sum_{\bm{k}} T_{V_{\mathrm{s}} \bm{k}} \neq 0$.
Moreover, $\sum_{\bm{k}} T_{\bm{k}}^{\mathrm{WTT}} \neq 0$.

We here further decompose the quartic-energy transfers.
Let us introduce the triad interaction functions 
corresponding to Eqs.~(\ref{eq:transferK4}) and (\ref{eq:transferVs})
as
\begin{subequations}
\begin{align}
 T_{K \bm{k}\bm{k}_1\bm{k}_2}^{(4)} 
&= \frac{|\bm{k}_1 \times \bm{k}_2|^2}{2\rho}
 p_{\bm{k}} (\zeta_{\bm{k}_1} \chi_{\bm{k}_2} + \chi_{\bm{k}_1} \zeta_{\bm{k}_2})
 \delta_{\bm{k}+\bm{k}_1+\bm{k}_2,\bm{0}}
\nonumber\\
&
\qquad
+ \mathrm{c.c.}
,
\\
T_{V_{\mathrm{s}} \bm{k}\bm{k}_1\bm{k}_2}
&=-\frac{|\bm{k}_1 \times \bm{k}_2|^2}{2\rho}
  \chi_{\bm{k}} (p_{\bm{k}_1} \zeta_{\bm{k}_2}  + \zeta_{\bm{k}_1} p_{\bm{k}_2})
 \delta_{\bm{k}+\bm{k}_1+\bm{k}_2,\bm{0}}
\nonumber\\
&
\qquad
+ \mathrm{c.c.}
,
\end{align}%
\end{subequations}%
which represent the transfer of each energy to $\bm{k}$ due to a triad
with one leg $\bm{k}_1$ and the other $\bm{k}_2$.
To symmetrize the triad interaction functions
and to make a triad in the form $\bm{k}+\bm{k}_1+\bm{k}_2=\bm{0}$,
we use $\zeta_{\bm{k}} = \zeta_{-\bm{k}}^{\ast}$, $p_{\bm{k}} = p_{-\bm{k}}^{\ast}$ and $\chi_{\bm{k}} = \chi_{-\bm{k}}^{\ast}$.
Then, the quartic-energy transfers can be represented as the sum of these terms:
\begin{align}
T_{K \bm{k}}^{(4)} 
=  \sum_{\bm{k}_1,\bm{k}_2}  T_{K \bm{k}\bm{k}_1\bm{k}_2}^{(4)} 
,
\quad
T_{V_{\mathrm{s}} \bm{k}} 
=\sum_{\bm{k}_1,\bm{k}_2} T_{V_{\mathrm{s}} \bm{k}\bm{k}_1\bm{k}_2}
.
\end{align}
The triad interaction function of the total energy is defined as
$T_{\bm{k}\bm{k}_1\bm{k}_2}=T_{K \bm{k}\bm{k}_1\bm{k}_2}^{(4)}+T_{V_{\mathrm{s}} \bm{k}\bm{k}_1\bm{k}_2}$.
The triad interaction function $T_{\bm{k}\bm{k}_1\bm{k}_2}$ is interpreted as 
the temporal rate of the energy increment at $\bm{k}$
due to the interaction among the three wavenumbers $\bm{k}+\bm{k}_1+\bm{k}_2=\bm{0}$.
The triad interaction function of the total energy satisfies the detailed energy balance:
\begin{align}
T_{\bm{k}\bm{k}_1\bm{k}_2}+T_{\bm{k}_1\bm{k}_2\bm{k}}+T_{\bm{k}_2\bm{k}\bm{k}_1}=0.
\label{eq:detailedbalance}
\end{align}
Namely,
the triad interaction function shows the interchanges of the energy among wavenumbers
keeping the sum of the energies of the three wavenumbers.

The triad interaction functions have high symmetry.
If we define the triad interaction functions in a piecewise way as
\begin{subequations}
\begin{align}
 \widetilde{T}_{K \bm{k}\bm{k}_1\bm{k}_2}^{(4)} 
&= \frac{|\bm{k}_1 \times \bm{k}_2|^2}{2\rho}
 p_{\bm{k}} \zeta_{\bm{k}_1} \chi_{\bm{k}_2}
 \delta_{\bm{k}+\bm{k}_1+\bm{k}_2,\bm{0}}
,
\\
\widetilde{T}_{V_{\mathrm{s}} \bm{k}\bm{k}_1\bm{k}_2}
&=-\frac{|\bm{k}_1 \times \bm{k}_2|^2}{2\rho}
  \chi_{\bm{k}} p_{\bm{k}_1} \zeta_{\bm{k}_2}
 \delta_{\bm{k}+\bm{k}_1+\bm{k}_2,\bm{0}}
,
\end{align}%
\label{eq:piecewisetriadfunc}%
\end{subequations}%
then another detailed energy balance holds:
\begin{align}
\widetilde{T}_{K \bm{k}\bm{k}_1\bm{k}_2}^{(4)} + \widetilde{T}_{V_{\mathrm{s}} \bm{k}_2 \bm{k} \bm{k}_1} = 0
.
\end{align}
This represents the gain of the kinetic energy at $\bm{k}$
and that of stretching energy at $\bm{k}_2$
have the same absolute value with the opposite signs
through the triad interaction atomized as Eqs.~(\ref{eq:piecewisetriadfunc}).
It indicates
the exchange between the kinetic energy and the stretching energy
through the triad interaction.
The atomized triad interaction function of the total energy is then defined as
\begin{align}
\widetilde{T}_{\bm{k}\bm{k}_1\bm{k}_2} &=
 \widetilde{T}_{K \bm{k}\bm{k}_1\bm{k}_2}^{(4)} + \widetilde{T}_{V_{\mathrm{s}} \bm{k}\bm{k}_1\bm{k}_2}
,
\end{align}%
and the detailed energy balance that is the same as Eq.~(\ref{eq:detailedbalance})
holds also for $\widetilde{T}_{\bm{k}\bm{k}_1\bm{k}_2}$.

The detailed energy balances hold via the triad interaction functions
among the Fourier coefficients of the physical variables,
$\zeta_{\bm{k}}$, $p_{\bm{k}}$ and $\chi_{\bm{k}}$.
It suggests that the present representation by using these Fourier coefficients
is suitable for the analysis of energy budget.
Since $\chi_{\bm{k}}$ is given by the convolution as defined in Eq.~(\ref{eq:chik}),
it is consistent with the fact that the nonlinear interactions occur among four waves
when the complex amplitudes are used for the governing equation~(\ref{eq:fvka}).

The azimuthally-integrated energy transfers,
which are defined in the similar way to the energy spectra,
are drawn in Fig.~\ref{fig:transfer}.
The azimuthally-integrated energy transfer $\mathcal{T}(k)$, for example, is 
defined as
$\mathcal{T}(k) = (\Delta k)^{-1} \sum_{k-\Delta k/2 \leq |\bm{k}^{\prime}| < k+\Delta k/2} \langle T_{\bm{k}^{\prime}} \rangle$.
The area between the solid red curve and the zero line
at the small wavenumbers equals to that at the large wavenumbers
which is enlarged in the inset.
It is consistent with the energy conservation.
The much larger amplitude of $\mathcal{T}(k)$ at the small wavenumbers
than that at the large wavenumbers
results from the logarithmically-scaled horizontal axis.
The dissipation scale can be estimated roughly as $k \approx 10^3$,
since the total-energy transfer $\mathcal{T}$ becomes large and positive in the wavenumbers
$256\pi \lessapprox k \lessapprox 512\pi$.
It is also consistent with the exponential decay of the energy spectra 
in Fig.~\ref{fig:alles}.

Because the quadratic transfers, or transmutations, can be rewritten as 
$T_{K \bm{k}}^{(2)} = - T_{V_{\mathrm{b}} \bm{k}} 
= - \omega_{\bm{k}}^2 \mathrm{Im}(a_{\bm{k}} a_{-\bm{k}})$, 
the result $\mathrm{Im}(C_{\mathrm{C} \bm{k}}) \approx 0$ 
shown in Fig.~\ref{fig:correlation}
is equivalent to the fact that 
both $\mathcal{T}_{K}^{(2)}(k)$ and $\mathcal{T}_{V_{\mathrm{b}}}(k)$ are almost $0$
over all the wavenumbers.
Namely,
the relatively large values of $\mathcal{T}_{K}^{(2)}(k)$ and $\mathcal{T}_{V_{\mathrm{b}}}(k)$
observed at the large wavenumbers
are caused by the statistical fluctuations,
and they diminish as the number of realizations increases.

Energy is transferred among wavenumbers by the quartic parts,
i.e., $\mathcal{T}_{K}^{(4)}(k)$ and $\mathcal{T}_{V_{\mathrm{s}}}(k)$.
Note that $\mathcal{T}^{\mathrm{WTT}}(k) = \mathcal{T}_{K}^{(4)}(k)$.
At the small wavenumbers
the transfer of the stretching energy $\mathcal{T}_{V_{\mathrm{s}}}(k)$
is dominant in that of the total energy $\mathcal{T}(k)$
and is negative.
In this region, the energy excited by the external force
is carried to the inertial subrange by $\mathcal{T}_{V_{\mathrm{s}}}(k)$.
In the statistically-steady state,
the negative energy transfer is canceled by the input due to the external force.
All the energy transfers are close to $0$ in the inertial subrange.
On the other hand,
at the large wavenumbers,
the quartic-energy transfer of the kinetic energy $\mathcal{T}_{K}^{(4)}(k)$
accounts for most of $\mathcal{T}(k)$,
and is positive.
The positive energy transfer is canceled by the output due to the dissipation.
Namely,
the wave field receives energies as the stretching energy from the external force,
the kinetic and stretching energies are transferred to the small scales,
and the wave field dissipates energies through the kinetic energy.

\begin{figure}
 \includegraphics[scale=1.05]{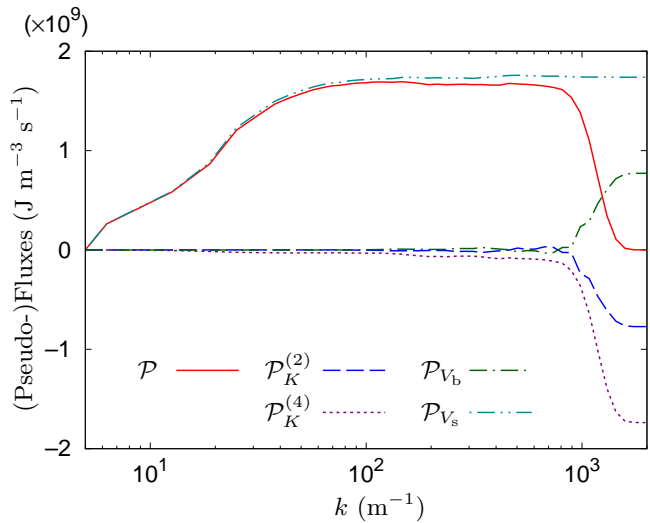}
 \caption{(Color online)
 Flux of the total energy,
 and pseudo-fluxes of the quadratic and quartic parts of the kinetic energy,
 and of the bending and stretching energies.
 \label{fig:flux}
}
\end{figure}

The conservation of the total energy leads to
the continuity of energy in the wavenumber space:
\begin{align}
\frac{\hat{d} E_{\bm{k}}}{\hat{d} t} + \nabla_{\bm{k}}\cdot {\bm{P}}_{\bm{k}}
= T_{\bm{k}} + \nabla_{\bm{k}}\cdot {\bm{P}}_{\bm{k}}
=0
.
\label{eq:defflux2D}
\end{align}
Here, $\bm{P}_{\bm{k}}$ is the two-dimensional flux of the total energy,
and $\nabla_{\bm{k}}\cdot$ is the divergence operator in the wavenumber space.
The locality of the energy cascade due to the nonlinear interactions is assumed.
In the statistically-isotropic system,
the continuity equation of the energy is
given by the azimuthal integration of Eq.~(\ref{eq:defflux2D})
as
\begin{align}
\mathcal{T}(k)
+ \frac{\partial \mathcal{P}(k)}{\partial k} = 0
.
\label{eq:defflux}
\end{align}
Then,
the total-energy flux $\mathcal{P}(k)$ can be represented by using the total-energy transfer
and has the indefiniteness of constants of the integration.
When we set the flux to be $0$ at the smallest wavenumber,
the flux is defined as
\begin{align}
\mathcal{P}(k) \equiv - \int_0^k \mathcal{T}(k^{\prime}) dk^{\prime}
.
\end{align}
The energy conservation guarantees $\mathcal{P}(\infty)=0$,
which allows us to rewrite the flux as
$\mathcal{P}(k) = \int_k^{\infty} \mathcal{T}(k^{\prime}) dk^{\prime}$.

On the other hand,
the flux of the quadratic energy is simply defined 
in terms of $\mathcal{T}^{\mathrm{WTT}}$
as
\begin{align*}
\mathcal{P}^{\mathrm{WTT}}(k) \equiv - \int_0^k \mathcal{T}^{\mathrm{WTT}}(k^{\prime}) dk^{\prime}
.
\end{align*}
However,
this quantity does not represent the flux of the quadratic energy,
since
the quadratic energy is not conserved.
Because the flux of the quadratic energy is ill-defined,
we here refer it as ``pseudo-flux'' of the quadratic energy.
The non-conservation of the quadratic energy 
results in
$\mathcal{P}^{\mathrm{WTT}}(k) \neq \int_k^{\infty} \mathcal{T}^{\mathrm{WTT}}(k^{\prime}) dk^{\prime}$,
and both of $\mathcal{P}^{\mathrm{WTT}}(0)$ and $\mathcal{P}^{\mathrm{WTT}}(\infty)$ cannot be $0$ at the same time.

In WTT,
the quadratic energy is conserved under the kinetic equation.
Therefore,
$\mathcal{P}^{\mathrm{WTT}}$
is physically-meaningful only in the weakly nonlinear limit.
However,
this can be extended to neither the finite nonlinearity
nor Hamiltonian systems which generally consist of both resonant and non-resonant terms.
In the earlier studies, nonetheless,
$\mathcal{P}^{\mathrm{WTT}}$
has been considered as the energy flux in weak turbulence,
while, in fact,
only the total-energy flux $\mathcal{P}$ is physically-meaningful.

The pseudo-flux of the quadratic energy in WTT is equal to 
the quartic part of the pseudo-flux of the kinetic energy.
To break the energy flux $\mathcal{P}$ into elements,
we forcibly define the pseudo-fluxes of the decomposed energies
in Eq.~(\ref{eq:energytransfers}) as
\begin{align*}
\mathcal{P}_i(k) \equiv - \int_0^k \mathcal{T}_i(k^{\prime}) dk^{\prime}
,
\end{align*}
similarly to $\mathcal{P}^{\mathrm{WTT}}$.
It is cautioned again that
we cannot expect the conservation of the decomposed energies
and the above definition is merely an expedience for comparison with the earlier studies.

The total-energy flux and these pseudo-fluxes are drawn in Fig.~\ref{fig:flux}.
The total-energy flux $\mathcal{P}$ is $0$ at the maximal wavenumber,
while the pseudo-fluxes are not $0$ there.
The non-zero value of $\mathcal{P}^{\mathrm{WTT}}=\mathcal{P}^{(4)}_K$ 
at the maximal wavenumber
results from the non-conservation of the quadratic energy,
and it was observed also in MMT model~\cite{Rumpf2005188}.
Furthermore, the value of $\mathcal{P}^{\mathrm{WTT}}=\mathcal{P}^{(4)}_K$ 
in the inertial subrange seems to be slightly negative,
which is opposite to that of the true energy flux $\mathcal{P}$,
though, of course, its sign as well as its value depends on the boundary condition for $\mathcal{P}^{\mathrm{WTT}}$. 

The weakly nonlinear spectrum and the strongly nonlinear spectrum
are respectively observed
at the large wavenumbers and at the small wavenumbers
in Fig.~\ref{fig:alles}.
In spite of the coexistence of the weakly and strongly nonlinear regimes,
the total-energy flux $\mathcal{P}$ is
almost constant in the inertial subrange by definition in the statistically-steady state.
The total-energy flux that is constant and positive in the inertial subrange
indicates the forward energy cascade.
One may naively predict that the energy flux $\mathcal{P}$ changes
according the flexion of $\mathcal{E}(k)$
or at the intersection of $\mathcal{V}_{\mathrm{b}}(k)$ and $\mathcal{V}_{\mathrm{s}}(k)$
in Fig.~\ref{fig:alles}.
Moreover, one might expect that
the large energy flux and the small energy flux
are respectively observed
at the strongly-nonlinear small wavenumbers and at the weakly-nonlinear large wavenumbers.
However, in fact, the energy flux in the statistically-steady state
is constant in the inertial subrange,
where neither the external force nor the dissipation affects it.

\section{Concluding remark}

In this paper,
the energy is decomposed into 
the kinetic, bending and stretching energies
in the elastic-wave turbulence governed by the F\"{o}ppl-von K\'{a}rm\'{a}n (FvK) equation.
The Fourier coefficient of the Airy stress potential appropriately gives 
the nonlinear energy, i.e., the stretching energy,
for a single wavenumber in the elastic waves.
The complex amplitude $a_{\bm{k}}$ has been introduced as an elementary wave to apply the random phase approximation
in researches of weak turbulence.
In fact,
$a_{\bm{k}}$ has clear physical meaning in analogy with the wave action,
and gives the sophisticated formalism in the weak turbulence theory (WTT).
However,
the use of the Fourier coefficients of physical variables, $\zeta_{\bm{k}}$, $p_{\bm{k}}$ and $\chi_{\bm{k}}$, is natural
for evaluation of energy,
since the nonlinear energy expressed by the complex amplitude $a_{\bm{k}}$
is given by the convolution.

By the energy decomposition analysis,
it was found that
the kinetic energy and the stretching energy 
are much larger than the bending energy
in the (relatively) strongly nonlinear regime,
while the bending energy comparable with the kinetic energy is much larger than the stretching energy
in the weakly nonlinear regime.
The imbalance between the kinetic and bending energies
results from the strong correlation between $a_{\bm{k}}$ and $a_{-\bm{k}}$.
In fact,
$a_{\bm{k}} \approx -a_{-\bm{k}}^{\ast}$ 
in the strongly nonlinear regime.
Although one may expect a distinctive structure in the real space
due to this correlation,
it is not so easy to identify it
because of the cumulative effect of all active modes.
Namely, the summation of the all active modes including phase correlation
makes the real-space structure.
It is our future work to clarify such properties.

The S-theory is developed
to explain the strong pairing between $a_{\bm{k}}$ and $a_{-\bm{k}}$
in the spin waves under strong parametric excitation~\cite{zakharov1971stationary}.
In this case,
the interactions among pairs are more essential 
than those among elementary waves.
The external force in the present study is not parametric,
though the pairing plays an important role in the strongly nonlinear regime.
Independently of the S-theory,
the pairing itself might be essential for the energy budget,
because not $a_{\bm{k}}$ but $\zeta_{\bm{k}}$, $p_{\bm{k}}$ and $\chi_{\bm{k}}$ are the basic elements,
and the nonlinear terms appear as $(a_{\bm{k}} + a_{-\bm{k}}^{\ast})$ in the governing equation.

As a result of the single-wavenumber representation of the nonlinear energy,
the analytical expression of the energy budget
was obtained for the first time in wave turbulence systems.
The quadratic-energy transfers,
which are the quadratic part of the kinetic-energy transfer
and the bending-energy transfer,
transmute the energies for a wavenumber.
Since the quartic part of the kinetic-energy transfer and the stretching-energy transfer
are the same quartic order of the complex amplitude,
both energy transfers should not be discriminated even in the weakly nonlinear limit.
The analytical expression of the energy budget shows that
the total-energy transfer,
which is sum of the quartic-energy transfers,
satisfies the detailed energy balance.
These facts indicate that the stretching energy has equal essentiality
to the kinetic energy
in considering the energy budget,
though the order of the stretching energy ($O(|a|^4)$)
is higher than that of the kinetic energy ($O(|a|^2)$)
in the complex-amplitude representation.
It was numerically found in the present system that
the energy is input into the system through the stretching-energy transfer
at small wavenumbers,
and dissipated through the quartic part of the kinetic-energy transfer
at large wavenumbers.

The energy transfer is defined as the rate of change of the energy,
and it holds independently from the total-energy conservation.
On the other hand,
the energy flux is defined based on the continuity equation of energy.
Therefore,
while the decomposed-energy transfer can reflect the energy budget,
the decomposed-energy flux cannot.
It follows that only the total-energy flux is the actual flux.
It is indispensable to include the nonlinear energy properly
to satisfy the energy conservation and to obtain the total-energy flux.
In order to compare with previous researches,
we introduced and examined the pseudo-fluxes as well,
though they are not actual but spurious fluxes,
since the conservation of energy 
which the fluxes rely on
does not hold for each decomposed energy.

We have succeeded for the first time to evaluate
the well-defined total-energy flux
directly by using the analytical expression of the total-energy transfer due to the nonlinear interactions.
The total-energy flux evaluated by the nonlinear terms
is positively constant in the inertial subrange,
and it indicates the forward energy cascade.
The fluxes of the quadratic energies reported in various wave turbulent systems~\cite{Rumpf2005188,pushkarev,*zakharov_steady,*PhysRevLett.96.204501}
have physical meaning only in the weakly nonlinear limit.
Because the external force used in Ref.~\cite{PhysRevE.89.062925} directly excites only
the linear energy, which is the kinetic energy,
the expression of the cumulative energy input $\widetilde{\mathcal{F}}(k)$
is indistinguishable from the one where the nonlinear energy is not considered.
This approach conceals the energy budget in the inertial subrange,
and loses the distinction between the quadratic and quartic energies.
For a general external force
that may excite the nonlinear energy directly,
the stretching-energy transfer should be taken into account
as pointed out above in the present paper.
Note that $\widetilde{\mathcal{F}}(k) - \widetilde{\mathcal{D}}(k)$,
which is used as a total-energy flux in the same reference~\cite{PhysRevE.89.062925},
is always constant in the inertial subrange
when both the external force and the dissipation are localized in the wavenumber space,
and hence the energy cascade cannot be examined by such flux.
The analytical expression of the energy flux obtained from the nonlinear terms in the governing equation
is necessary to investigate the wave turbulence statistics in the inertial subrange.

Although one may expect to evaluate the energy flux by using the expression 
based on the two-points structure functions in the real space
as usually done in analyses of hydrodynamic turbulence,
it may be difficult to evaluate those for the nonlinear energy in wave systems.
It is because the nonlinearity in such systems appears
as the higher-order expansion of the complex amplitudes
in contrast with the success of the K\'arm\'an-Howarth relation 
in the Navier-Stokes turbulence
where the total energy is represented in the quadratic form.
One might be able to find alternative ways to go beyond in this direction
by introducing adequate modes of physical quantities.

It is of interest that 
the total-energy fluxes are nearly equal in both weak and strong turbulence regimes
while the two regimes coexist in the inertial subrange.
It may show another mechanism than those considered in the critical balance,
e.g., turning of the energy transfer in quasi-geostrophic turbulence,
since the present system is statistically isotropic in contrast with those where the critical balance is predicted~\cite{nazarenkobook}.

\appendix
\section{Hamiltonian structure expressed in terms of complex amplitude}
\label{sec:Appendix}

The complex amplitude $a_{\bm{k}}$ introduced in Eq.~(\ref{eq:ComplexAmplitude})
plays a role as a canonical variable:
\begin{align*}
i\frac{d {a}_{\bm{k}}}{dt}=\frac{\delta \mathcal{H}}{\delta a^{\ast}_{\bm{k}}},
\end{align*}
because the Hamiltonian can be rewritten in terms of the complex amplitude as
\begin{align}
 \mathcal{H} =& \sum_{\bm{k}} \omega_{\bm{k}} |a_{\bm{k}}|^2
  + 
 \!\!\!\!\!\!\!\!
 \sum_{\bm{k}+\bm{k}_1-\bm{k}_2-\bm{k}_3=\bm{0}}
 \!\!\!\!\!\!\!\!
  W^{\bm{k}\bm{k}_1}_{\bm{k}_2\bm{k}_3}
  a_{\bm{k}} a_{\bm{k}_1} a_{\bm{k}_2}^{\ast} a_{\bm{k}_3}^{\ast}
 \nonumber\\
 &
  +
 \!\!\!\!\!\!\!\!
  \sum_{\bm{k} - \bm{k}_1 - \bm{k}_2 - \bm{k}_3 = \bm{0}}
 \!\!\!\!\!\!\!\!
 \left(
  G^{\bm{k}}_{\bm{k}_1\bm{k}_2\bm{k}_3}
  a_{\bm{k}} a_{\bm{k}_1}^{\ast} a_{\bm{k}_2}^{\ast} a_{\bm{k}_3}^{\ast}
 + \mathrm{c.c.}
 \right)
 \nonumber\\
 &
  + 
 \!\!\!\!\!\!\!\!
 \sum_{\bm{k}+\bm{k}_1+\bm{k}_2+\bm{k}_3=\bm{0}}
 \!\!\!\!\!\!\!\!
 \left(
  R_{\bm{k}\bm{k}_1\bm{k}_2\bm{k}_3}
  a_{\bm{k}} a_{\bm{k}_1} a_{\bm{k}_2} a_{\bm{k}_3}
 + \mathrm{c.c.}
 \right)
 .
\label{eq:Hak}
\end{align}
The second, third, and fourth terms respectively show 
the $2 \leftrightarrow 2$, $1 \leftrightarrow 3$ and $0 \leftrightarrow 4$ interactions 
of the four-wave interactions,
and $W^{\bm{k}\bm{k}_1}_{\bm{k}_2\bm{k}_3} (=W^{\bm{k}_2\bm{k}_3\ast}_{\bm{k}\bm{k}_1})$,
$G^{\bm{k}}_{\bm{k}_1\bm{k}_2\bm{k}_3}$ and
$R_{\bm{k}\bm{k}_1\bm{k}_2\bm{k}_3}$ are the matrix elements of the interactions.
Note that the interactions include both resonant and non-resonant interactions.
Only under the kinetic equation of WTT,
where only the resonant terms are retained,
the quadratic energy is conserved. 

The third and fourth terms of the Hamiltonian~(\ref{eq:Hak}) 
are rarely taken into account in the literature~\cite{zak_book},
because these terms can often be reduced by a canonical transformation
in the weak turbulence regime of most wave turbulence systems~\cite{kras}.
In the elastic-wave turbulence,
the fourth term can be reduced,
but the third term cannot be as known from the linear dispersion relation~(\ref{eq:lineardispersion}),
which allows the $1 \leftrightarrow 3$ resonant interactions. 
The $1 \leftrightarrow 3$ interactions of the Hamiltonian
results in the $1 \leftrightarrow 3$ resonant interactions in the kinetic equation.
It indicates that the wave action is not conserved
even according to its kinetic equation in WTT.
The existence of the $1 \leftrightarrow 3$ resonant interactions
is one of the distinctive feature of the present system \cite{during2006weak}.

\begin{acknowledgments}
Numerical computation in this work was carried out at the Yukawa Institute Computer Facility.
This work was partially supported by KAKENHI Grant No.~25400412.
\end{acknowledgments}

\end{document}